\documentclass[preprint,pre,floats,aps,amsmath,amssymb]{revtex4}

\usepackage{graphicx}
\usepackage{bm}
\usepackage{comment}

\setcitestyle{super}

\begin{document}

\title{5-Methylation of Cytosine in CG:CG Base-pair Steps: \\ 
A Physicochemical Mechanism for the Epigenetic \\
Control of DNA Nanomechanics}
\author{Tahir I. Yusufaly$^{1}$, Yun Li$^{2}$ and Wilma K. Olson$^{3*}$}
\affiliation{[1] Rutgers, the State University of New Jersey, Department of Physics and Astronomy, Piscataway, NJ, USA, 08854 \\
                   [2] Delaware Valley College, Department of Chemistry and Biochemistry, Doylestown, PA, USA, 18901 \\
		[3] Rutgers, the State University of New Jersey, Department of Chemistry and Chemical Biology, Piscataway, NJ, USA, 08854 \\		
		* Corresponding Author. Email: wilma.olson@rutgers.edu; Phone: 732-445-4619 }
\date{\today}

\begin{abstract}
Van der Waals density functional theory is integrated with analysis of a non-redundant set of protein-DNA crystal structures from the Nucleic Acid Database to study the stacking energetics of CG:CG base-pair steps, specifically the role of cytosine 5-methylation. Principal component analysis of the steps reveals the dominant collective motions to correspond to a tensile `opening' mode and two shear `sliding' and `tearing' modes in the orthogonal plane. The stacking interactions of the methyl groups globally inhibit CG:CG step overtwisting while simultaneously softening the modes locally via potential energy modulations that create metastable states. Additionally, the indirect effects of the methyl groups on possible base-pair steps neighboring CG:CG are observed to be of comparable importance to their direct effects on CG:CG. The results have implications for the epigenetic control of DNA mechanics. 
\end{abstract}

\maketitle

\section{Introduction}

Self-consistent Kohn-Sham density functional theory (KS-DFT) \cite{HK64,KS65} has traditionally been one of the most popular tools of choice for ab initio electronic structure calculations of properties of dense matter, due to both its comparable accuracy to quantum chemical methods and relatively small computational cost. However, until recently, the failure of traditional exchange-correlation functionals to account for nonlocal London dispersion forces precluded its application to sparse matter, such as biological molecules. The development of the nonlocal van der Waals density functional vdW-DF \cite{Dion04} and the subsequent, higher accuracy vdW-DF2 \cite{Lee10} has expanded the realm of DFT to a whole new class of systems.

\begin{figure}[b]
   \begin{center}
      \includegraphics[width=3.25in]{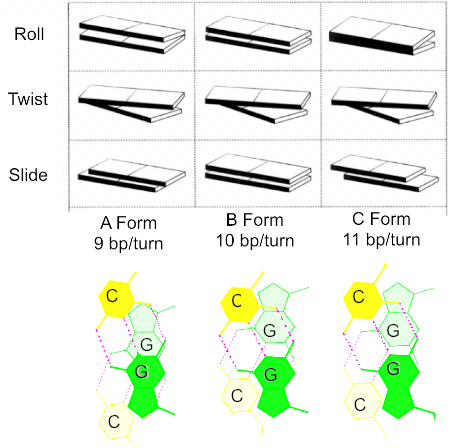}
      \caption{Schematic illustration of important base-pair step parameters of canonical A, B and C forms of DNA. Also included are sample stacking diagrams for a CG:CG step, showing the exact spatial displacements of chemical units in fiber models \cite{Arnott99}. The lower rigid body in the upper schematics is denoted by the lightly shaded base pair in the lower stacking diagram. The shaded edges on the schematic blocks and the right edges of the stacking diagrams both correspond to the minor-groove edges of base pairs. The pink dashed lines represent hydrogen bonds. Schematics adapted from Reference 6 and stacked CG step images computed with X3DNA \cite{ZhengLuOlson09}.}
     \label{ABCDNA}
   \end{center}
\end{figure}

One of the first applications of vdW-DF was the analysis by Cooper et al. \cite{Cooper08} of hydrogen bonding between nucleic-acid base pairs and stacking interactions between successive bases in base-pair steps. The authors successfully accounted for sequence-specific trends in base-pair separation and rotation seen in high-resolution crystal structures \cite{GhorinZurkinOlson95}. Additionally, they demonstrated the role that the methyl group of thymine plays in stabilizing double-stranded DNA over its uracil counterpart in RNA. The initial success of vdW-DF, combined with the ensuing development of the even more accurate vdW-DF2, motivates deeper theoretical study of the structural energetics of biologically relevant nucleobase configurations. 

In vivo, DNA predominantly adopts right-handed double-helical structures. Therefore, this class of structures is likely to be the most biologically relevant. In particular, high-resolution DNA crystal structures primarily adopt three right-handed double-helical states, namely A, BI and BII. Repetitions of these local conformations throughout the nucleic acid lead to polymers characterized as A-like, B-like and C-like, respectively \cite{OlsonSrini09, Arnott99, VanDamLevitt00}. 

The subtle differences in base-pair step geometries for each of the different helical states are illustrated in Figure \ref{ABCDNA}. Most notably, A-like base-pair steps are under-twisted relative to B-like ones, while C-like steps are over-twisted. It has been suggested \cite{Tolstorukov07} that the interconversions between these three configurations are related to sequence-specific mechanisms of nucleosome positioning in chromatin, the bundled assembly of DNA and proteins in eukaryotic cell nuclei. This has consequences for the control of gene silencing.

\begin{figure}
   \begin{center}
      \includegraphics[width=3.15in]{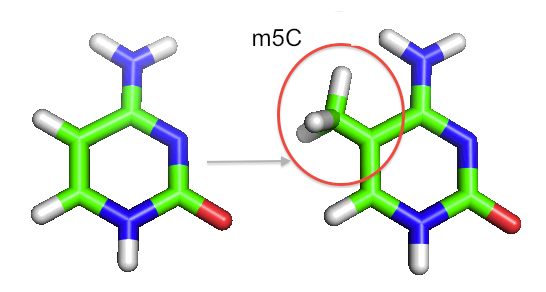}
      \caption{5-methylation of cytosine, an important epigenetic modification. The methyl group that replaces the hydrogen is circled in red. Green represents carbon, white hydrogen, red oxygen, and blue nitrogen. Graphics generated using PyMOL \cite{Pymol}.}
     \label{5Methylation}
   \end{center}
\end{figure}

The stabilizing effect of methylation on thymine found by Cooper et al. inspires the question of whether similar effects occur for other biochemical modifications. Specifically, the methylation of cytosine at the C5 position, as illustrated in Figure \ref{5Methylation}, is an important epigenetic modification. In particular, in CpG dinucleotides, it appears to trigger the protein-assisted compaction of chromatin \cite{BirdWolfe99}. This compaction determines whether or not genes can be transcribed into RNA. Elucidating the effects of the 5-methylation of cytosine on the structural energetics of CG:CG steps is therefore a valuable step towards understanding the factors that control eukaryotic gene regulation. Also worth analyzing are possible indirect effects, in particular, the interactions of methylated cytosines of CG:CG steps with immediately adjacent base-pair steps, including AC:GT, TC:GA, CC:GG and GC:GC.

In addition, there has been very little, if any, work connecting first-principles electronic structure calculations of biomolecular structure with bioinformatics analyses. The crystal structures in the Nucleic-Acid Database (NDB) \cite{BermanOlson92} can be utilized to extract the most commonly observed sequence-specific collective atomic motions. This allows reduction of the complex conformational coordinate space to a simpler subspace that is more likely to be biologically relevant. This information, besides being valuable in its own right, can interface with DFT calculations to determine realistic energy landscapes of stacked base pairs.  

Previous quantum chemical studies have focused on base stacking \cite{Sponer, Parker}, including some recent work on the stacking of Watson-Crick paired bases \cite{SponerNew}, and the effects of cytosine methylation in the context of reaction kinetics and equilibrium structures \cite{Forde, Youngblood, Yang}. However, these methods have not traditionally been integrated with a bioinformatics approach to look at DNA conformational energetics in the context of the coordinated motions of neighboring base pairs in the double helix. In this study, the analysis of Cooper et al. is extended to the 5-methylation of cytosine, taking into account these considerations. 

\begin{figure}[b]
   \begin{center}
      \includegraphics[width=5.5in]{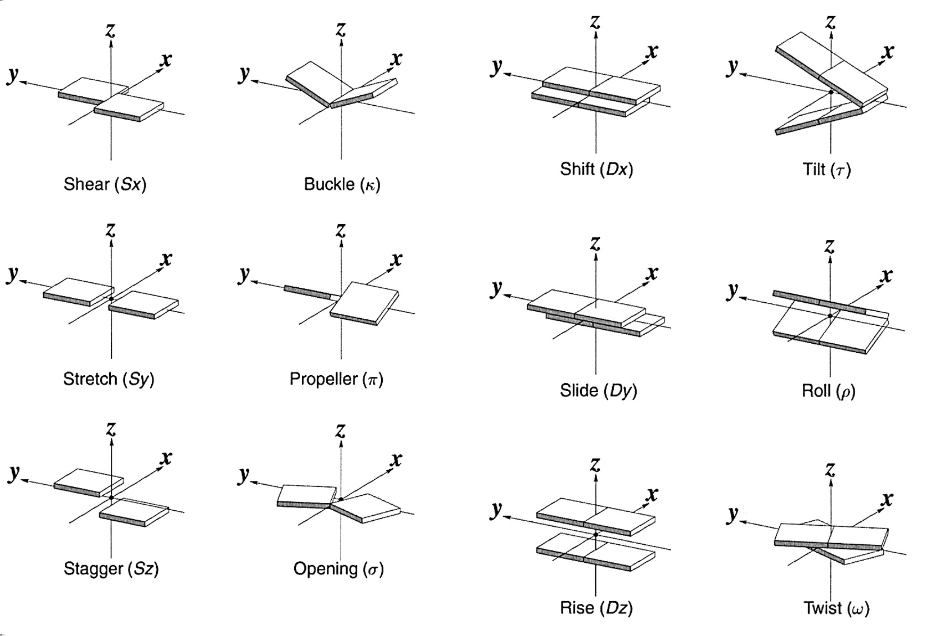}
      \caption{The rigid-body configuration of a DNA base-pair step is specified by six local parameters per base pair and six step parameters for successive base-pair steps. There are three translational and three rotational degrees of freedom for each kind of rigid-body motion. Figure adapted with permission from Reference 23.}
     \label{BPStepsAndRigids}
   \end{center}
\end{figure}

\section{Methods}

\subsection{Generation of a Non-Redundant DNA-Protein Dataset}
The motions of neighboring base-pair steps were deduced from a non-redundant dataset of 239 protein-DNA crystal complexes of 2.5 \AA{} or better resolution taken from the NDB \cite{Li}. The dataset included 101 structures of double-helical DNA bound to enzymes, 121 duplexes determined in the presence of regulatory proteins, 16 complexes with structural proteins, and one DNA bound to a multi-functional protein. The structures were filtered to exclude over-represented complexes in order to obtain a balanced sample of spatial and functional forms. The selection and classification of structures was based on sequential and structural alignment, as well as available protein classification databases, including the SCOP \cite{SCOP} scheme. The working dataset excluded terminal base-pair steps (i.e., residues at chain ends and nicked dimer steps), which may adopt alternate conformations or be affected by crystal packing, as well as chemically modified nucleotides, nucleotides involved in non-canonical base pairs, and ÔmeltedÕ residues, in which complementary base pairs are highly distorted and do not contain the requisite number or types of hydrogen bonds.

\subsection{Principal Component Analysis}
Information on dinucleotide steps was extracted from the NDB files using the 3DNA software package \cite{ZhengLuOlson09}. From this information, an eighteen-parameter data vector characterizing the conformation of a given base-pair step was generated with 3DNA. This included twelve base-pair parameters (six for each base-pair) and six step parameters (Figure \ref{BPStepsAndRigids}). After converting the data vector to a standardized z-value, principal component analysis was performed for the CG:CG steps. Using a scree test, the three highest eigenvalues, corresponding to the dominant collective modes of motion, were extracted. This process was repeated for all possible steps that could directly connect to CG:CG, and therefore be affected by methylation. 

\subsection{Density Functional Theory Calculations}
In the analysis of individual modes, a sufficient set of points within two standard deviations of the mean along the modal pathways was sampled to determine the shape of the energy landscape of stacked base pairs. Base-pair steps were created with 3DNA and methyl groups were added with Open Babel \cite{OBabel11}. Computations were performed using vdW-DF2 as implemented in the Quantum Espresso package \cite{Giannozzi09} via the algorithm of Roman-Perez and Soler \cite{RomanPerez09}. All calculations used standard generalized gradient approximation pseudopotentials \cite{TroullierMartins93}, with an energy cutoff of 60 Ry (1 Ry = 313.755 kcal/mol). SCF diagonalizations were performed with the Davidson algorithm, using convergence criteria of $10^{-6}$ Ry. To reduce spurious interaction between periodic images, the system was placed in a 40 x 30 x 30 cubic Bohr (1 Bohr = 0.529 Angstrom) supercell.

\section{Results and Discussion}

\begin{figure}[b]
   \begin{center}
      \includegraphics[width=5.0in]{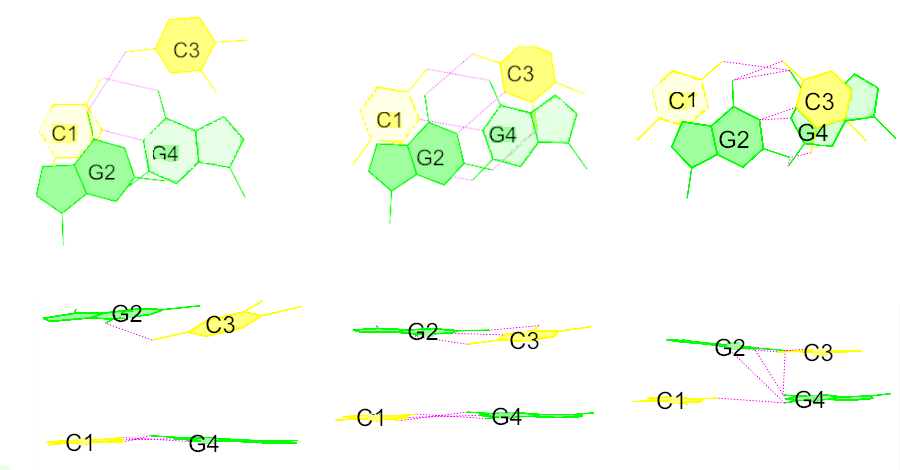}
      \caption{The first principal component of CG:CG steps, a tensile `opening' mode of the crack between DNA strands. From left to right, respectively, are images for steps that are five negative normal mode units from the mean, at the mean, and five positive units away. The upper and lower rows display views from the top-down and looking into the minor groove. The lower base pair is labelled by C1 bonded to G4, while the upper one is labelled by G2 and C3. The pink dashed lines represent hydrogen bonds. Molecular images created with 3DNA \cite{ZhengLuOlson09}.}
     \label{CGCGPCA1}
   \end{center}
\end{figure}

\begin{figure}[h]
   \begin{center}
      \includegraphics[width=5.0in]{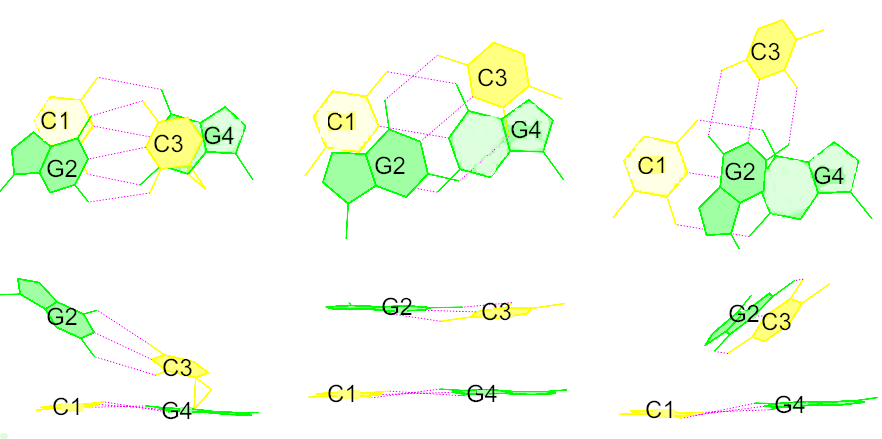}
      \caption{The second principal component of CG:CG steps, a shear `sliding' mode. See the caption of Figure \ref{CGCGPCA1} for explanation of notations and symbols.}
     \label{CGCGPCA2}
   \end{center}
\end{figure}

\begin{figure}[t]
   \begin{center}
      \includegraphics[width=5.0in]{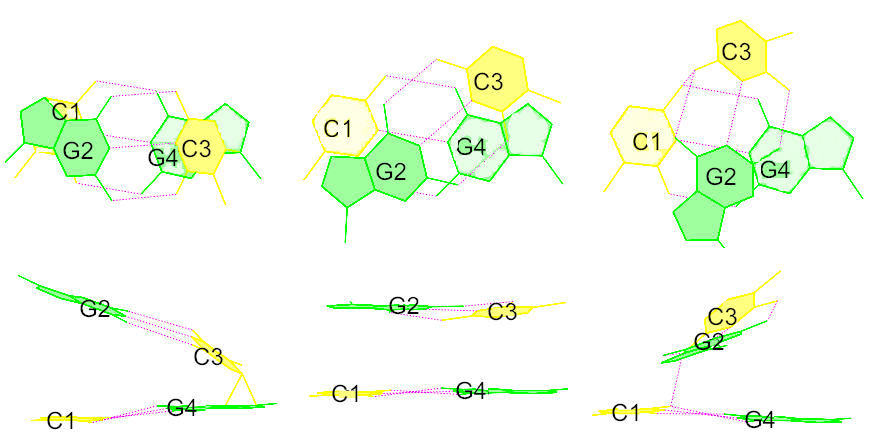}
      \caption{The third principal component of CG:CG steps, a shear `tearing' mode. See the caption of Figure \ref{CGCGPCA1} for explanation of notations and symbols.}
     \label{CGCGPCA3}
   \end{center}
\end{figure}

The main focus of this paper regards the effects of methylation on CG:CG steps. The next subsection discusses the qualitative nature of the modal motion of the three dominant principal components, which are termed `Opening', `Sliding' and 'Tearing' and illustrated in Figures \ref{CGCGPCA1}, \ref{CGCGPCA2} and \ref{CGCGPCA3}, respectively. This is followed up with a report on the effects of 5-methylcytosine on the stacking energetics of CG:CG steps, including the interplay between local and global effects. Finally, the indirect effects on neighboring steps are discussed.

\subsection{Nature of the Principal Components}
   
%The average values and standard deviations of the parameters are listed in Table \ref{table:CGCGParams}, along with the parameter motions along the three dominant principal component axes. Particularly significant parameter motions are highlighted in bold. 

The predominant principal component of CG:CG steps may be interpreted, borrowing a term from the fracture mechanics community \cite{LiuChaoZhu}, as a tensile `opening' mode between the two DNA strands. It consists of a coherent twisting, sliding and rising of the step, accompanied by out-of-phase stretching and opening of the C:G and G:C base pairs. As the vertical separation between the base pairs decreases and the step is tightened and undertwisted, the lower C:G pair breaks apart and the upper G:C pair is compacted. The most prominent consequence is an enhanced overlap between the C3 and G4 bases. 

This breakage of hydrogen bonds is similar to the `breathing' modes observed in the classic work of Mandal, Kallenbach and Englander \cite{EnglanderKallenbach}. The analysis here provides a detailed picture of the possible nature of this breaking, including the symmetric way in which the bond-breaking acts with respect to the equilibrium state.

The two secondary modes share several similarities. Both of them at their extreme ends correspond to the C and G bases along one strand having a very small separation, with the bases on the opposite strand being consequently more spread out and isolated. Motion along the modes can be interpreted as the two strands alternating between compressing and opening. In both cases, this is primarily controlled by a twisting motion that acts in the opposite direction of tilt and slide, and in tandem with shift. However, the remaining step parameters, namely rise and roll, move in an opposite direction in one of the modes relative to the other, as do dominant local base-pair motions, such as stagger or buckle. Thus, as the modes are traversed, the DNA gradually transitions from the A to B to C-forms.

To continue with the analogy with fracture mechanics, whereas the most dominant principal component represents a tensile `opening' mode, the two secondary components may be interpreted as an orthogonal pair of shearing modes. Together, they span the fracture plane perpendicular to that of the tensile opening. The conventional terms for this pair \cite{LiuChaoZhu} are `tearing' and `sliding' modes, respectively, and this nomenclature shall be adopted for the remainder of the paper. 

While the `opening' modes correspond to traditional breathing modes, this pair of transverse modes is more along the lines of hydrogen bond bifurcations \cite{Nelson87}. Here, the bases on neighboring stacks can become `mixed', resulting in hydrogen bonding not just between complementary bases in a single base pair, but also between adjacent ones located on the same DNA strand. 

\subsection{Effects of Cytosine 5-Methylation on CG:CG Steps}

\begin{figure}[t]
   \begin{center}
      \includegraphics[width=5.3in]{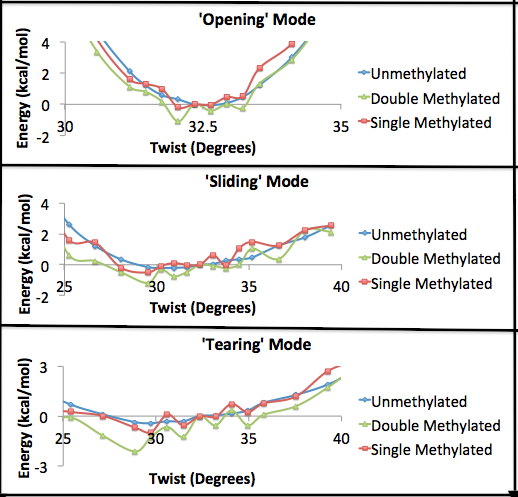}
      \caption{Stacking energy landscapes for each of the three principal components, with and without methylation. The horizontal axes are labelled by the variation of twist along each of the modes.}
     \label{CGCG}
   \end{center}
\end{figure}

\begin{figure}[h]
   \begin{center}
      \includegraphics[width=5.3in]{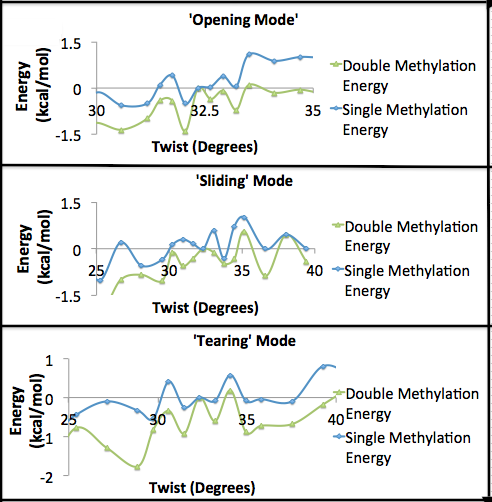}
      \caption{The individual contributions of one and two methyl groups to the effective stacking potential of the step, as measured by the energy difference, with respect to the unmethylated state, of calculations with one or both C5 groups methylated, respectively. The horizontal axes are labelled by the variation of twist along each of the modes.}
     \label{CGCGMethylation}
   \end{center}
\end{figure}

\begin{figure}[h]
   \begin{center}
       \includegraphics[width=5.0in]{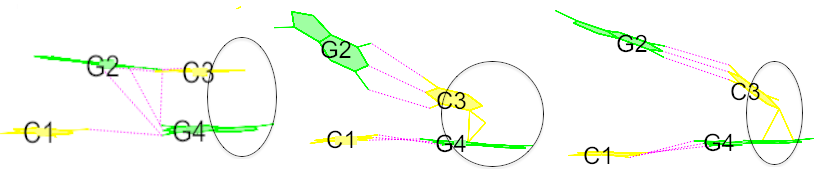}
      \caption{From left to right are minor-groove views of the low-twist regimes of the opening, sliding and tearing modes, respectively. As illustrated, in these regimes, the overlap area between C3 and G4 is greater. This leads to an enhancement in the stacking interactions of a methyl group at the C5 position with the adjacent guanine. The pink dashed lines represent hydrogen bonds.}
     \label{undertwisting}
   \end{center}
\end{figure}

The stacking energetics and effects of methylation for each of the principal modes are illustrated in Figures \ref{CGCG} and \ref{CGCGMethylation}, respectively. For convenience of illustration, the data are plotted against the value of the twist angle along each of the three mode landscapes. Two general trends emerge from the data: 1) The methyl groups globally suppress the overtwisting of CG:CG steps by pushing the minimum to a lower twist, and 2) This global inhibition is accompanied by local modulations of the potential that soften the landscape by creating an ensemble of intermediate states, including several that are overtwisted. 

To demonstrate the physical reason for the global inhibition of twisting, it is informative to look at the `low-twist' regimes of each of the modes. As illustrated in Figure \ref{undertwisting}, these regimes correspond to a higher degree of stacking overlap between the C3 and G4 bases. As a consequence, the stacking interactions of the C5 carbon in cytosine with the adjacent guanine are enhanced. The addition of a methyl group amplifies these stacking forces, serving as an effective `pinning' field that stabilizes undertwisted configurations.

\begin{figure}[t]
   \begin{center}
       \includegraphics[width=3.0in]{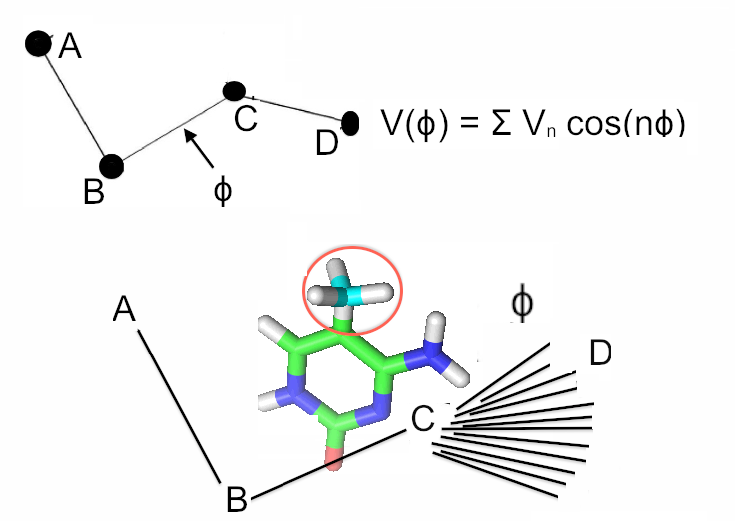}
      \caption{As a base-pair step moves along the landscape of its conformational modes, the methyl group may be interpreted as having a set of moving `non-covalent' dihedral angles with the atoms in the step, generated by the `angle' that the 5-methylcytosine makes with the remainder of the base-pair step. In analogy to its more commonly discussed analog in covalent bonding, this torsional variation creates an effective torsional `potential', which, like any periodic potential, can be decomposed into a superposition of harmonics of varying wavelength. In this schematic, A may be interpreted as a methyl group, the B-C line as the cytosine, and D as all other atoms in the step.}
     \label{NonLocalTorsion}
   \end{center}
\end{figure}

The origin of the local fluctuations is more subtle. A useful metaphor for illustrating where they come from, at least on a phenomenological level, is to consider the dihedral angle $\phi$ between four atoms in a covalent bond, as shown in Figure \ref{NonLocalTorsion}. As is well known \cite{Wales}, due to the periodicity of the dihedral angle, $\phi = \phi + 2 \pi$, the potential energy $V(\phi)$ can be expanded in a harmonic series. Often, one harmonic dominates the torsional energetics, but generally the potential is an incoherent superposition of different frequencies, leading to a complex landscape of metastable vacua. 

From the perspective of the CG:CG steps, the fundamental difference upon addition of the methyl groups is the introduction of several additional variables, which can be interpreted as a set of non-local `torsional' angles describing the orientation of the methyl group with respect to other atoms in the polymer assembly. These variables, being periodic like dihedral angles, likewise give rise to modulations in the potential energy. The net result of all such contributions is a noisy signal that `softens' the base-pair step, enabling it to more easily change its shape via fluctuations. This ties in nicely with the expectation that the 5-methylation of cytosine would make the base-pair softer, similar to previous observations by Cooper et al. regarding the impact that the methyl group of thymine has on A:T base pairs compared to A:U \cite{Cooper08}.

\subsection{Effects of Methylation on Neighboring Steps}
\begin{figure}[b]
   \begin{center}
     \includegraphics[width=7.0in]{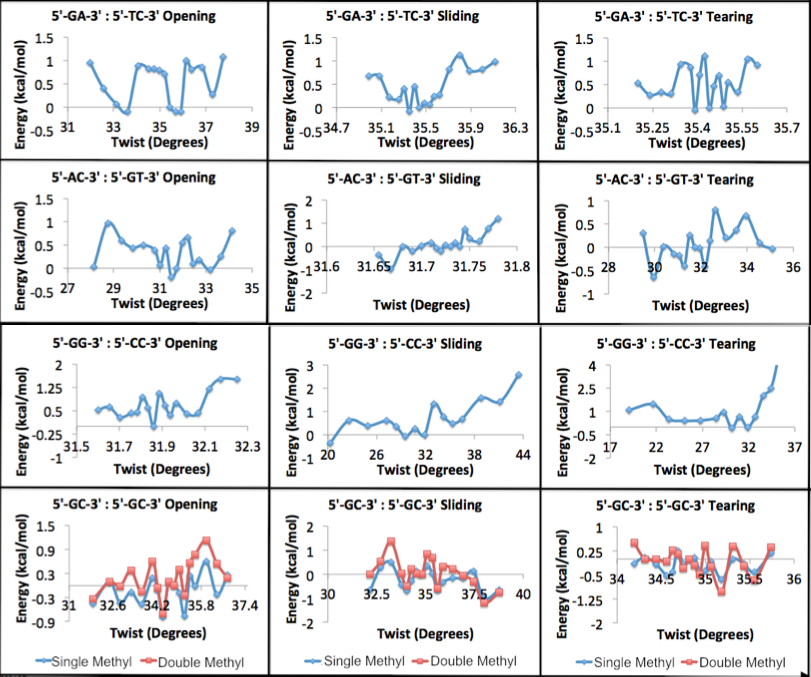}
      \caption{Indirect stacking interaction energies of 5-methylcytosine with possible steps neighboring the CG:CG step. For GA:TC, AC:GT, and GG:CC steps, there is only one possible cytosine that can connect to a CG:CG, and thus be potentially methylated. For GC:GC steps, however, it is possible for one or both of its cytosines to be methylated.}
     \label{MethylationEnergy}
     \end{center}
\end{figure}
Calculations of the energetics of different base-pair steps adjacent to CG:CG indicate that the interactions of the C5 methyl groups with the neighboring steps have an energetic effect comparable to that of their immediate interactions with the CG:CG step.

Principal component analysis reveals that the classification of dominant collective motions into opening, sliding and tearing modes persists for the different base-pair steps. Furthermore, as illustrated in Figure \ref{MethylationEnergy}, 5-methylcytosine continues to torsionally modulate the stacking energy as the identities of the remaining bases in the step change. 

However, there are also some important differences compared to CG:CG. Perhaps most relevant to nucleosome formation is the observation that 5-methylcytosine globally \textit{enhances} overtwisting of several modes in GC:GC steps. This overwinding is a potential mechanism for preserving the double-helical structure of DNA by countering the tendency to melt from CG:CG unwinding. 

Additionally, the typical stacking energy of methylation is lower for CG:CG steps than it is for its neighbors, as seen by a comparison of Figures \ref{CGCGMethylation} and \ref{MethylationEnergy}. From a statistical mechanical point of view, methylation enhances the room-temperature Boltzmann partition function of CG:CG steps while decreasing that of its neighbors, corresponding to an increase or decrease in Helmholtz free energy, respectively. Consequently, the methylation of CG:CG steps is more thermodynamically stable than methylation of other possible steps, an argument for why it is more commonly observed. %Write about sequence specificity, free energy stabilization of CGCG vs destabilization of others (why this one is favored), and potential for overtwisted states in latter...
% Indirect stacking interaction energies of 5-methylcytosine with possible steps neighboring a CG:CG. As in the main analysis of the paper, the methyl groups can be interpreted as effectively applying external non-local torsional potentials to the normal base-pair step. Among the consequences of these external fields are the potential stabilization of a number of local metastable states, including several that are overtwisted.

\section{Conclusions}
In summary, this study has extended the work of Cooper et al. in a systematic study of the effects of C5 methylation on base-stacking energetics. Methylation is seen to have nontrivial effects on on the flexibilities of the opening, sliding and tearing motions of CG:CG steps. Specifically, it globally inhibits overtwisted states while simultaneously generating local potential energy modulations that soften the step. Furthermore, analysis of interactions of the methyl groups with possible neighboring steps indicates that these effects are of comparable importance to those of the methyl group on CG:CG itself.

The mechanisms discussed in this work do not appear to be limited to this specific system.  There is consistent evidence that the methyl groups perform a functional role via a combination of long-wavelength and short-wavelength effects, which is suggestive of some more general principles underlying chemical epigenetic modifications and the physical processes responsible for their biological functionality, particularly in a mechanical context. 

The results of this work compare favorably with previous experimental data regarding the effects of cytosine methylation on nucleosome positioning. In particular, Davey, Pennings and Allan \cite{Davey1997} observed that methylation of nucleosomal DNA prevents the histone octamer from interacting with an otherwise high-affinity chicken $\beta$-globin gene positioning sequence. This sequence contains a (CpG)$_{3}$ motif located 1.5 helical turns from the dyad axis of the nucleosome, with minor-groove edges on the base-pair step that are oriented towards the histone core. When this sequence motif is unmethylated, it is capable of adopting the structural deformations necessary to interact with the histone octamer, and thus enable nucleosome positioning. However, as the current calculations demonstrate, methylation of CG-rich stretches of DNA enhances the formation of the A-DNA polymorph, a helical form that is more resistant to bending deformations than B-DNA, and which also bends DNA in the opposite sense. Consequently, interactions with the histones are inhibited, and nucleosome formation is suppressed. 

Furthermore, a followup study by Davey, et al. \cite{Davey2004} indicated that mutations of the (CG)$_{3}$ sequence motif into either GC:GC or CC:GG base-pair steps affect both the degree of nucleosome formation and the amount of disruption by CG:CG methylation. This ties in with the present finding that the effects of methylation depend on the sequential and structural context of the modified cytosines.

%The Supplementary Information reports results regarding potential indirect interactions of the CG:CG methyl groups with adjacent base-pair steps, including the change in the room-temperature Boltzmann partition function due to the free energy of methylation, which softens CG:CG steps while stiffening their neighbors.

This work, additionally, demonstrates a foundation for future studies of realistic structural biomaterials modeling at the atomistic level via density functional theory. In particular, the consistency between experiments and calculations, in both this work and in the earlier studies of Cooper et al. \cite{Cooper08}, points to the capability of using first-principles approaches to extract valuable biochemical information on systems in which there is no prior experimental data. Thus, density functional theory calculations can serve as a complement to more traditional single-molecule biophysical experiments.  

\section*{Acknowledgments}
We acknowledge valuable discussions with Kyuho Lee, Timo Thonhauser and Valentino Cooper regarding the use of vdW-DFT. Bohdan Schneider and Andrew Colasanti advised on the selection of structures for the determination of rigid-body parameters. T.Y. thanks Bell Laboratories for financial support through the Lucent Fellowship. This work was partially funded by the U.S.P.H.S. through GM 34809. 

\section*{Supporting Information Available}
The supporting information contains numerical output of the analysis of the three dominant principal components for each of the base-pair steps, including information on component coefficients and latent scores. Additionally, it contains comprehensive information on the non-redundant set of protein-DNA crystal complexes used in this study, including references to the original literature. This information is available free of charge via the Internet at http://pubs.acs.org/.

\begin{figure}[h]
   \begin{center}
       \includegraphics[width=3.0in]{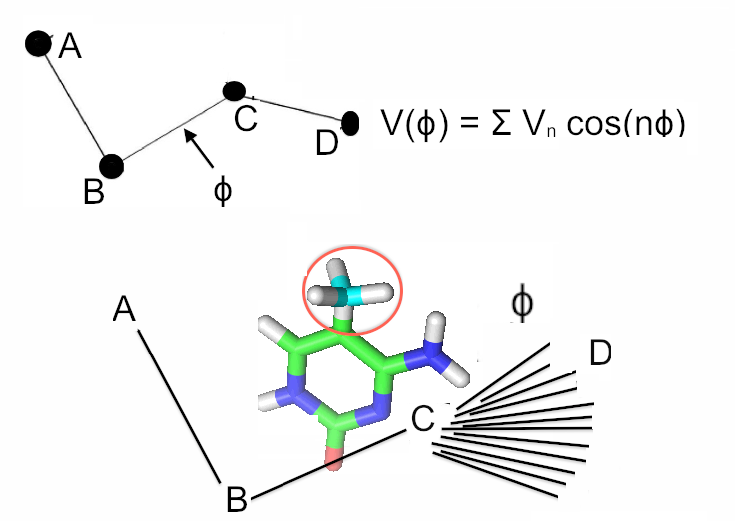}
      \caption{Table of Contents Figure}
     \label{TOC}
   \end{center}
\end{figure}

\end{document}